\documentstyle[12pt]{article}
\begin{document}
\large
\begin{center}
{\bf The Distinctive Feature of Weak Interactions and Some
of Its Subsequences (Impossibility of Generation of Masses and Absence of the MSW Effect)}\\
\vspace{0.2cm}
\par
Kh.M. Beshtoev
\vspace{0.2cm}
\par
Joint Institute for Nuclear Research, Joliot Curie 6
141980 Dubna, Moscow region, Russia; Scientific Research Institute of Applied
Mathematics and Automation of the Kabardino-Balkarian Scientific Center of
RAC, Shortanova 89a, 360017 Nalchik, Russia
\vspace{0.2cm}
\end{center}
\par
{\bf Abstract}
\par
In the Quantum theory the wave functions form a full and orthonormalized
functional space. For this reason we can use the equation on eigenfunctions
and eigenstates and find the eigenenergies $E_n$ and eigenfunctions $\Psi_n$
to determine the physical characteristics of the considered systems
(or models). In the Quantum theory the observed values are the average
value of operators. Since the wave functions create a full and orthonormalized
space, the average values of operators coincide with eigenvalues of operators.
This situation takes place in the case of strong and electromagnetic
interactions. However, the average values of the weak interaction operators
are equal to zero since only the left-handed components of
spinors participate in
the weak interactions. It means that in these interactions the connected states
cannot exist. Then the weak interaction of the particles
(scatterings, decays) can be considered only in the framework of the
standard perturbative approach.  It is shown that the weak interactions
cannot generate masses and the equation for Green's
function of the weak interacting fermions (neutrinos) in the matter
coincides with the equation for Green's function of fermions in vacuum.
And in the result we come to a conclusion:  the mechanism of resonance
enhancement of neutrino oscillations in matter (i.e.  MSW effect)
cannot exist.
\par
\noindent
PACS: 12.15 Ff-Quark and lepton masses and mixings.\\
PACS: 96.40 Tv-Neutrinos and muons.
\newpage
\section{Introduction}

In the strong and electromagnetic interactions the left-handed and
right-handed components of spinors participate in a symmetric manner. In
contrast to these interactions only the left-handed
components of spinors participate in the weak interactions.
This is a distinctive feature of the weak interactions.
\par
In three different approaches: by using mass Lagrangian [1, 2], by using
the Dirac equation [3, 2], and using the operator formalism [4],
I discussed the problem
of the mass generation in the standard weak interactions. The result was:
the standard weak interaction cannot generate masses of fermions since
the right-handed components of fermions do not participate in these
interactions. Then using this result in works [4] it has been shown that
the effect of resonance enhancement of neutrino oscillations in matter
cannot exist.
\par
At present there is a number of papers published (see [5] and references
there) where by using the Green's function
method it is shown that the weak interactions can generate the
resonance enhancement of neutrino oscillations in matter (it means that
the weak interaction can generate masses).
As we see below, this result is a consequence of using the
weak interaction term $H^{int}_\mu = V_\mu \frac{1}{2}(1- \gamma_5)$ in
an incorrect manner, and
in the result they have obtained that the right-handed components of the
fermions participate in the weak interactions.
\par
Now let us come to a common consideration and then consider concrete
examples and consequences of the distinctive feature of the weak interactions.

\section{\bf Common Consideration Consequences of Distinctive Feature of
Weak Interactions}

In the Quantum theory the wave functions form a full and orthonormalized
functional space. For this reason we can use the equation on eigenfunctions
and eigenvalues
$$
\hat E \Psi(...) = F(...) \Psi(...) = 0 ,
\eqno(1)
$$
and find the eigenenergies $E_n$ and eigenfunctions $\Psi_n$
to determine the physical characteristics of the considered systems [6, 7]
(or models). In the Quantum theory the observed values are the average
value of operators
$$
E_n = (\Psi_n, \hat E \Psi_n)  .
\eqno(2)
$$
Indeed, since the wave functions create a full and orthonormalized space,
the average values of operators coincide with the eigenvalues of operators.
This situation takes place in the case of strong and electromagnetic
interactions and in concrete objects which appear from these interactions.
It takes place since the wave functions and their conjunction function exist
in this case. An absolutely another situation takes place in the case of the
weak interactions. In these interactions only the left-handed components of
the wave function (wave vector) (i.e. the left-handed components of
spinors) participate in
these interactions. In the weak interaction the $P$-symmetry is violated.
And what is more in the theory for his $\gamma_5$ invariance, it is more
suitable to use the Dirac [8] or Veil [9] but not the Shr${\ddot{o}}$dinger
type of equations.  Naturally, the following question arises: To which
consequences does this distinctive feature of the weak interactions lead?
\par
Also in the strong and electromagnetic interactions in the weak interactions
we can use the perturbative theory but in this case propagators must be
propagators of free particles (without renormalization).
\par
Let $\bar \Psi_L, \Psi_L, \bar \Psi_R, \Psi_R$-be wave functions (wave vectors)
of spinor particles. Since we consider the weak interactions where
the left-handed components of spinors participate, then
$$
\bar \Psi_R = \Psi_R \equiv 0, \qquad \bar \Psi_L = \bar \Psi \frac{1}{2}
(1 + \gamma_5) ,  \Psi_L = \frac{1}{2}(1 - \gamma_5) \Psi .
\eqno(3)
$$
If $\hat B$ is an operator of the weak interactions, then
$$
\hat B \Psi_L = B \Psi_L ,
\eqno(4)
$$
and then the mean value of this operator (the observed value) is
$$
\bar B = (\bar\Psi, \hat B\Psi_L) = B (\bar \Psi_R, \Psi_L) = 0 .
\eqno(5)
$$
It is interesting to see: which values are zero in the weak interactions?
\par
Now consider the problem of eigenstates and eigenvalues in the weak
interactions. Let $\hat F$ be an operator and we divide it into two parts.
The first part $\hat A$ characterizes the free particle, and the second
part $\hat B$ is responsible for the weak interaction, then
$$
\hat F = \hat A + \hat B ,
$$
$$
\hat F\Psi = \hat A\Psi + \hat B\Psi ,
\eqno(6)
$$
and the mean value of $\hat F$ is
$$
(\bar\Psi, \hat F\Psi) = (\bar\Psi, \hat A\Psi) +
(\bar\Psi_R, \hat B\Psi_L) + (\bar\Psi_L, \hat B\Psi_R) =
$$
$$
(\bar\Psi, \hat A\Psi) + (\bar \Psi_R \equiv 0)(\bar\Psi_R, \hat B\Psi_L)
+ (\Psi_R \equiv 0)(\bar\Psi_L, \hat B\Psi_R)
= (\bar\Psi, \hat A\Psi) .
\eqno(7)
$$
The obtained result means that in the weak interactions there cannot arise
the connected
states in contrast to the strong and electromagnetic interactions. Besides,
the average value of the polarization operators  is equal to zero, i.e. the
polarization of the matter is absent. In the same way we can show that
the equation for
renormcharge for the weak interaction is equivalent to the equation for the free
charge, i.e. renormcharge $Q^2(t)$ in the weak interactions [10] (where $t$ is
a transfer momentum squared) does not change and $Q^2(t) = const$ in
contrast to renormcharges $e^2(t), g^2(t)$ of the electromagnetic and strong
interactions [11] (it is necessary to remark that the neutral current of the
weak interactions
includes a left-right symmetrical part which is renormalized).
\par
Let us consider the equation for Green's function of
fermions taking into account the standard weak interactions.

\section{Equation for Green's Function in Weak Interactions}

The Green's function method is frequently used for taking into account
effects of electromagnetic interactions and strong interactions
(chromodynamics) [12]. The equation for Green's function has the following
form:
$$
[\gamma^\mu (i \partial_\mu - V_\mu) - M] G(x, y) = \delta^4 (x - y) ,
\eqno(8)
$$
where $V_\mu$ characterizes the electromagnetic or strong interactions and
$$
i G(x, y) = <T \Psi(x) \bar \Psi(y)>_o .
$$
It is necessary to mention that the Green's function method is a very
convenient
method for studying the electromagnetic and strong interaction effects
since these interactions are left-right side symmetrical interactions.
\par
At present a  number of works has been published where Green's function
was used for taking into account the weak interaction. There was shown that
the weak interaction can generate masses, i.e. masses of fermions are changed
by the weak interactions and then the resonance enhancement of neutrino
oscillations appears in the matter [5]. In this work we want to show that
this result is a consequence that there the distinctive feature of the
standard weak interactions is not used correctly, namely, that the right-handed
components of fermions do not participate in these interactions (i.e. $\Psi_R =
\bar \Psi_R \equiv 0$).
\par
Usually the equation for Green's function for fermion (neutrino) with weak
interactions  is taken in the following form:
$$
[\gamma^\mu (i \partial_\mu - V_\mu) - M] G(x, y) = \delta^4 (x - y) ,
\eqno(9)
$$
where $V_\mu$ is
$$
V_\mu = V_\mu \frac{1}{2}(1 - \gamma_5) = V_\mu P_L .
\eqno(10)
$$
\par
It is supposed that the term (10) in Eq.(9) reproduces the distinctive
feature of the weak interactions
$$
V_\mu G(x, y) \to V_\mu \frac{1}{4}(1 - \gamma_5)^2 T(\Psi(x) \bar \Psi(y))
 = V_\mu T(\Psi_L(x) \bar \Psi_R(y)) .
$$
However, this operation is not correct since it does
not reproduce the standard weak interaction.  We see that, if we
directly use the distinctive feature of these interactions, then the
equation for Green's function must be rewritten in the form
$$
[\gamma^\mu (i \partial_\mu - V_\mu \left[\begin{array}{cc} \Psi_R = 0
\\ \bar \Psi_R = 0 \end{array}\right]) - M] G(x, y) = \delta^4 (x - y) .
\eqno(11)
$$
Then the interaction term in Eq.(11) is
$$
V_\mu \left[\begin{array}{cc}\Psi_R = 0 \\ \bar \Psi_R = 0 \end{array}\right]
T(\Psi_L \bar \Psi_R) =
$$
$$
 = T(\Psi_L \bar \Psi_R (\bar \Psi_R = 0) + (\Psi_R = 0) \Psi_R \bar \Psi_L)
 = V_\mu 0 \equiv 0 ,
\eqno(12)
$$
and then Eq.(11) is transformed in the following equation:
$$
[\gamma^\mu (i \partial_\mu) - M] G(x, y) = \delta^4 (x - y) ,
\eqno(13)
$$
which coincides with the equation for free Green's function (i.e. equation
without interactions). So, we see that
the equation for Green's function with weak interactions (in matter) coincides
with the equation for Green's function in vacuum.

\section{Impossibility to Realize  the Mechanism of Resonance
Enhancement of Neutrino Oscillations in Matter}

In the previous part we have obtained that the equation for Green's function
of fermions with weak interactions has the form (13). It is a consequence of
the fact that the right-handed components of fermions (neutrinos) do not
participate in the weak interactions. It means that the weak interaction
cannot generate masses (see also works [1-4]) and, correspondingly, the weak
interactions do not give a deposit to effective masses of fermions (neutrinos)
therefore, the mixing angle cannot be changed in weak interactions (in matter)
and it coincides with the mixing angle in vacuum.
\par
The two neutrino ($a, b$) mixing angle in vacuum is given by the expression
[13]
$$
sin^2 2\theta = \frac{(2 m_{\nu_a\nu_b})^2}{(m_{\nu_a}  - m_{\nu_b})^2 +
(2 m_{\nu_a\nu_b})^2}  ,
\eqno(14)
$$
and this mixing angle $\theta_m$ in the matter is
$$
sin^2 2\theta_m = \frac{(2 m_{\nu_a\nu_b})^2}{(m'_{\nu_a}  - m'_{\nu_b})^2 +
(2 m_{\nu_a\nu_b})^2}   ,
\eqno(15)
$$
where $m_{\nu_a}, m_{\nu_b}, m_{\nu_a \nu_b}$ are masses of neutrinos $a, b$,
nondiagonal mass term, and $m'_{\nu_a}, m'_{\nu_a}$-effective masses of the
same neutrinos in matter.
Since the masses of neutrinos $a, b$ in vacuum and in the matter
$$
m'_{\nu_a} = m_{\nu_a} \qquad m'_{\nu_b} = m_{\nu_b}
\eqno(16)
$$
are equal for the distinctive feature of the weak interactions, then the
mixing angles in vacuum $sin^2 2\theta$ and in the matter $sin^2 2\theta_m$
coincides. Hence, the mechanism of the
resonance enhancement of neutrino oscillations in the matter (MSW effect)
cannot exist.
\par
The problem of the resonance enhancement of neutrino oscillations in the matter
can be solved by another approach. Namely, while neutrino passing  through
the matter there can arise a polarization of the matter. Then
the problem is in computation of the energy of polarization of the matter.
In Wolfenstein's equation
for neutrinos [14] the term responsible for the (hypothetical) weak interaction
is exactly the energy of polarization [15] (it is worth reminding that
the energy and momentum transfer at elastic scattering in the forward direction
is zero). As it is stressed above, we can use the previous scheme for
computation of the polarization energy of matter by neutrinos in the weak
interactions. If $\epsilon$ is the operator for polarization energy of
matter by neutrinos, then the average value is
$$
\bar \epsilon = (\bar\psi_R, \hat \epsilon \Psi_L) =
\epsilon (\bar\Psi_R \equiv 0)(\Psi_R, \Psi_L) = 0 .
\eqno(17)
$$
We see that the Wolfenstein's equation for (real) neutrino in the matter
coincides with the equation for free neutrinos, then no resonance enhancement
of neutrino oscillations in the matter appears.

\section{Conclusion}

In the Quantum theory the wave functions form a full and orthonormalized
functional space. For this reason we can use the equation on eigenfunctions
and eigenstates and find the eigenenergies $E_n$ and eigenfunctions $\Psi_n$
to determine the physical characteristics of the considered systems
(or models). In the Quantum theory the observed values are the average
value of operators. Since the wave functions create a full and orthonormalized
space, the average values of operators coincide with eigenvalues of operators.
This situation takes place in the case of strong and electromagnetic
interactions. However, the average values of the weak interaction operators
are equal to zero, since only the left-handed components of
spinors participate in
the weak interactions. It means that the connected states
cannot exist in these interactions. Then the weak interaction of the
particles (scatterings, decays) can be considered only in the framework
of the standard perturbative approach.  It was shown that the weak
interactions cannot generate masses and the equation for Green's
function of the weak interacting fermions (neutrinos) in the matter
coincides with the equation for Green's function of fermions in vacuum.
And in the result we come to a conclusion:  the mechanism of resonance
enhancement of neutrino oscillations in matter (i.e.  MSW effect)
cannot exist.
\par
In conclusion I would like to
stress that in the experimental data from [16] there is no visible change in
the spectrum of the $B^{8}$ Sun neutrinos. The measured spectrum of
neutrinos lies lower than the computed spectrum of the $B^{8}$
neutrinos [17]. In the case of realization of the resonance
enhancement  mechanism this spectrum must be distorted. Also, the
day-night effect on the neutrinos regeneration in bulk of the Earth
is preserved within the mistakes [16], i.e. it is not observed.

\par
References
\par
\noindent
1. Kh.M. Beshtoev, JINR Communication P2-93-44, Dubna, 1993.
\par
\noindent
2. Kh.M. Beshtoev, Fiz. Elem. Chastitz At. Yadra (Phys.
\par
Part. Nucl.), {\bf 27}, 53(1996).
\par
\noindent
3. Kh.M.  Beshtoev, JINR Communication E2-93-167, Dubna, 1993.
\par
\noindent
4. Kh. M. Beshtoev, HEP-PH/9912532, 1999;
\par
Hadronic Journal, {\bf 22}, 235(1999).
\par
\noindent
5. C.Y. Cardall, D.J. Chung, Phys. Rev. {\bf D60}, 073012(1999).
\par
\noindent
6. V. Pauli, Principles of Wave Mechanics, OGIS, Moscow, 1947;
\par
P. Dirac, Principles of Quantum Mechanics, Oxford,
\par
Clarendon Press, 1958.
\par
\noindent
7. S. S. Schweber. An Introduction to Relat. Quantum Field Theory,
Row, Peterson and Co., N.Y., 1961.
\par
\noindent
8. Particles and Fields, Phys. Rev. {\bf D45}, N11(1998);
\par
L.B. Okun, Leptons and Quarks, M., Nauka, 1990.
\par
\noindent
9. S.S. Schweber, An Introduction to Relativistic Quantum
\par
Field Theory, Row, Peterson and Co., N. Y., 1961, p.114.

\par
\noindent
10. Kh.M. Beshtoev, JINR Communication E2-94-31, Dubna, 1994.
\par
\noindent
11. T. Izikson amd K. Suber, Quant. Field Theory, M., 1984;
\par
N.N. Bogolubov, D.V. Schirkov, Intr. to Theory of Quant.
\par
Field, M., 1994, p.469.
\par
\noindent
12. J. Schwinger, Phys. Rev. {\bf 76}, 790(1949);
\par
F.J. Dyson, Phys. Rev. {\bf 85}, 631(1952);
\par
N.N. Bogolubov, D.V. Schirkov, Intr. to Theory of Quant.
\par
Field, M., 1994, p.372, 437;
\par
Y. Nambu and G. Jona-Lasinio, Phys. Rev. {\bf 122}, 345(1961);
\par
M.K. Volkov, Fiz. Elem. Chastitz At. Yadra, {\bf 17}, 433(1986);
\par
A.A. Abrikosov et al., Methods of Quant. Theory Field in
\par
Statistic Phys. M., 1962.
\par
\noindent
13. Kh.M. Beshtoev, JINR Communication E2-99-307, Dubna, 1999;
\par
HEP-PH/9911513.
\par
\noindent
14. L. Wolfenstein, Phys. Rev. {\bf D14}, 2369(1978).
\par
\noindent
15. Kh.M. Beshtoev, Hadronic Journ., {\bf 22}, 477(1999).
\par
\noindent
16. K.S. Harita, et al., Phys. Rev. Lett. {\bf65}, 1297(1991);
\par
Phys. Rev., {\bf D44}, 2341(1991);
\par
Y. Totsuka, Rep. Prog. Physics 377, (1992).
\par
Y. Suzuki, Proceed. of the Intern. Conf. Neutrino-98, Japan, 1998.
\par
\noindent
17. J.N. Bahcall, Neutrino Astrophysics, Cambridge U.P.
\par
Cambridge, 1989.

\end{document}